\documentclass[reprint,aps,amsmath,amssymb,superscriptaddress,longbibliography,prx]{revtex4-2}
\usepackage{float}
\usepackage{placeins}
\usepackage{multirow}
\usepackage{booktabs}
\usepackage{comment}
\usepackage{graphicx}
\usepackage{dcolumn}
\usepackage{bm}
\usepackage{layouts}
\usepackage{url}
\usepackage[colorlinks=true, linkcolor=byzantine, citecolor=emerald, urlcolor=emerald, breaklinks=true]{hyperref}
\usepackage{color}
\usepackage{palatino}
\usepackage{newtxmath}
\usepackage{braket}
\usepackage{physics}
\usepackage[d]{esvect}
\usepackage{nicefrac}
\usepackage{tikz}
\usepackage{circuitikz}
\newcommand{\pder}[2][]{\frac{\partial#1}{\partial#2}}
\makeatletter
\def\@bibdataout@aps{%
\immediate\write\@bibdataout{%
@CONTROL{%
apsrev41Control%
\longbibliography@sw{%
    ,author="08",editor="1",pages="1",title="0",year="1"%
    }{%
    ,author="08",editor="1",pages="1",title="",year="1"%
    }%
  }%
}%
\if@filesw \immediate \write \@auxout {\string \citation {apsrev41Control}}\fi 
}
\makeatother

\begin{document}
\definecolor{byzantine}{rgb}{0.74, 0.2, 0.64}
\definecolor{emerald}{rgb}{0.25,0.5,0.27}
\title{
Geometric superinductance qubits: 
\\Controlling phase delocalization 
across a single Josephson junction
}
\author{Matilda Peruzzo}
\email{matilda.peruzzo@ist.ac.at}
\affiliation{Institute of Science and Technology Austria, 3400 Klosterneuburg, Austria}
\author{Farid Hassani}
\affiliation{Institute of Science and Technology Austria, 3400 Klosterneuburg, Austria}
\author{Gregory Szep}
\affiliation{King's College London, London WC2R 2LS, United Kingdom}
\author{Andrea Trioni}
\affiliation{Institute of Science and Technology Austria, 3400 Klosterneuburg, Austria}
\author{Elena Redchenko}
\affiliation{Institute of Science and Technology Austria, 3400 Klosterneuburg, Austria}
\author{Martin \v{Z}emli\v{c}ka}
\affiliation{Institute of Science and Technology Austria, 3400 Klosterneuburg, Austria}
\author{Johannes M.~Fink}
\email{jfink@ist.ac.at}
\affiliation{Institute of Science and Technology Austria, 3400 Klosterneuburg, Austria}

\date{\today}

\begin{abstract}
There are two elementary superconducting qubit types that derive directly from the quantum harmonic oscillator. In one the inductor is \emph{replaced} by a nonlinear Josephson junction to realize the widely used charge qubits with a compact phase variable and a discrete charge wavefunction. In the other the junction is \emph{added} in parallel, which gives rise to an extended phase variable, continuous wavefunctions and a rich energy level structure due to the loop topology.
While the corresponding rf-SQUID Hamiltonian was introduced as a quadratic, quasi-1D potential approximation to describe the fluxonium qubit implemented with long Josephson junction arrays, in this work we implement it directly using a linear superinductor formed by a single uninterrupted aluminum wire. We present a large variety of qubits all stemming from the same circuit but with drastically different characteristic energy scales. This includes flux and fluxonium qubits but also the recently introduced quasi-charge qubit with strongly enhanced zero point phase fluctuations and a heavily suppressed flux dispersion.
The use of a geometric inductor results in high precision of the inductive and capacitive energy as guaranteed by top-down lithography - a key ingredient for intrinsically protected superconducting qubits. The geometric fluxonium also exhibits a large magnetic dipole, which renders it an interesting new candidate for quantum sensing applications.
\end{abstract}

\maketitle








\section{Introduction}
Superconducting qubits are highly engineerable quantum systems that are at the forefront of quantum technology due to the offered strong interactions resulting in fast and precise control but also due to the similarity to existing microchip fabrication and the available diversity of circuit designs \cite{Clarke2008,Vool2017,Krantz2019}. Creative new ideas on how to encode, store and control single quanta in electrical circuits paired with state of the art fabrication have not only led to a big push in coherence times but also facilitated the observation of many new quantum physics phenomena \cite{Blais2021}.  

One example of this development is the fluxonium qubit \cite{Manucharyan2009}, a large inductance rf-SQUID characterized by strong anharmonicity and flux tunability. The ability to create such qubits with small transition frequency and charge matrix element around the flux sweet spot has led to 
the energy relaxation time $T_1$ being greatly extended \cite{Nguyen2019,Earnest2018,Somoroff2021} but driving the qubit transition also becomes increasingly difficult for the same reason. On the opposite side of the rf-SQUID spectrum the qubit can be made flux insensitive by increasing the shunting inductance due to enhanced quantum phase fluctuations. This bears resemblance to the transmon qubit where increasing the shunting capacitance decreases charge noise sensitivity via enhanced quantum charge fluctuations \cite{Koch2007}. This limit was recently explored in Ref.~\cite{Pechenezhskiy2020} using a suspended Josephson junction array 
with a ground state phase delocalization probability $|\varphi|>\pi$ of up to 40\% leading to a theoretically predicted flux-limited coherence time $T_\text{2}$ on the order of hundreds of $\mu$s. This ultra-high impedance limit is difficult to enter in practice, because it requires fabrication of a large inductor without creating a sizable capacitance that would inevitably lower the qubit's first transition frequency. The superinductance limit was therefore widely believed to be inaccessible with conventional inductors \cite{Manucharyan2009,Masluk2012} until very recently~\cite{Peruzzo2020}. 

\begin{figure*}[t]
\centering
\includegraphics[width=1.0\textwidth]{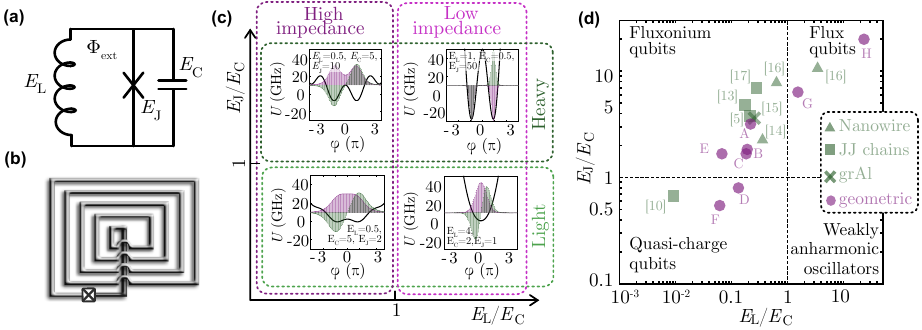}
\caption{\textbf{rf-SQUID qubit classification.}
(a) rf-SQUID circuit with inductive energy $E_{L}$, Josephson energy $E_{J}$, capacitive energy $E_{C}$ and external flux $\Phi_\textrm{ext}$. 
(b) Sketch of a geometric rf-SQUID qubit - an exact implementation of the circuit in panel (a).
(c) Classification scheme according to the ratios between the relevant energy scales at $\Phi_\text{ext} = 0.5\, \Phi_0$. Each quadrant shows the qubit potential $U$ as a function of phase $\varphi$ superimposed with the ground state (purple) and excited state (green) wavefunctions (scaled for visibility). 
Different qubit limits are determined by whether the qubit characteristic impedance $Z_C \propto \sqrt{E_{C}/E_{L}}$ is large or small compared to the resistance quantum $R_Q \approx 6.45$ k$\Omega$ and whether the qubit is heavy ($E_{J}/E_{C} > 1$, dominated by capacitance) or light ($E_{J}/E_{C} < 1$, dominated by tunneling). 
(d) Geometric rf-SQUID qubits presented in this paper (purple) alongside other qubits taken from the literature \cite{Manucharyan2009,Pop2014,Peltonen2018, Grunhaupt2019,Hazard2019,Pechenezhskiy2020,Zhang2021} (green) placed within the classification presented in panel (c). The four classes, identified by their limit cases 
are the fluxonium qubit, the flux qubit, the quasi-charge qubit and the weakly anharmonic oscillator limit. The extracted energies and design parameters for the geometric qubits are reported in Table \ref{tab:summary}.
} 
\label{fig:fluxonium_modelling}
\end{figure*} 

So far superinductors have been realized by means of kinetic inductance, either using disordered materials with strong electron scattering \cite{Grunhaupt2019} or by taking advantage of the inductance of a long chain of large Josephson junctions \cite{Masluk2012,Bell2012,Pechenezhskiy2020}. However, these techniques come with their own challenges. Josephson junction arrays and granular aluminum fragment the superconducting wavefunction, which leaves the qubit potentially affected by charge noise \cite{Mizel2020,DiPaolo2021} or disorder due to static charges \cite{Cedergren2017} that may also limit their applications in metrology \cite{Piquemal2000}. Josephson junction arrays are additionally subjected to critical current noise, which adds fluctuations to the value of the inductance \cite{Vool2017}. In the case of granular aluminum qubits there is a difference in critical temperature between the inductor and the junction 
that may lead the junction to attract quasiparticles 
and can give rise to a higher loss rate \cite{Grunhaupt2019}. 
In addition, a high degree of reproducibility of thin films and tunnel barriers is challenging due to their strong dependence on the exact growth and lithography conditions, a challenge that for single junction transmon devices has recently been addressed with individual laser assisted annealing \cite{Hertzberg2020}. 


In this work we introduce a wide range of single Josephson junction rf-SQUID qubits that rely on a geometric shunt inductor realized from a highly miniaturized $\sim 50\,\mu$m diameter planar aluminum coil suspended on a 220 nm thick silicon membrane - the device layer of a commercial silicon on insulator wafer. This new circuit element introduced in Ref.~\cite{Peruzzo2020} has low-loss, behaves linearly, is easy to design and reproduce and can be described as a lumped LC circuit up to the frequency region of its first self resonance. 
For a circular design with many turns $n\gg1$ and large filling factor the capacitive energy $E_C=e^2/(2C)$ is simply given by the coil's outer radius and the inductive energy $E_L=(\Phi_0/(2\pi))^2/L$ with $\Phi_\text{0}$ the magnetic flux quantum by how many turns are fit into that given radius. In contrast to previous implementations such a geometric inductor is a simple 2-terminal device that hosts a single uninterrupted superconducting wavefunction with a fixed and geometrically defined inductance and distributed capacitance offering a high degree of control and design flexibility. Qubits made from such linear inductors are expected to be less sensitive to quasiparticles owing to their small kinetic contribution ($<10$\,\% \cite{Peruzzo2020}) and because the lower disorder 
results in a shorter quasiparticle lifetime.

We show how the same physical circuit shown in Fig.~\ref{fig:fluxonium_modelling} (a) and (b) gives rise to different qubit types and qubit properties. 
Of the eight qubits studied, some reside in the classical 
flux qubit limit \cite{Yan2016} characterized by a strong localization of phase with zero point phase fluctuations of only $\varphi_\textrm{zpf}=(2E_C/E_L)^{1/4}=0.56$, while others reach the quasi-charge regime \cite{Pechenezhskiy2020} characterized by a strongly reduced flux sensitivity due to a wavefunction probability delocalization $|\varphi|>\pi$ of up to 30\% ($\varphi_\textrm{zpf}=2.4$) at half flux $(\Phi_\text{ext} =0.5\,\Phi_0$). 
This complements parallel work to delocalize the circuit ground state phase by means of Cooper pair co-tunneling \cite{Smith2020} and represents an important step towards realizing passively protected circuits \cite{Doucot2012, Dat2019} such as the 0-$\pi$ \cite{Brooks2013,Groszkowski2018,Gyenis2021} or the $\cos{(2\varphi)}$ \cite{SmithKou2020}. The qubit states of such devices would be spanned by two degenerate ground states protected by circuit symmetries and thus will rely on a very precise control of the qubit energies - properties that are ensured by top-down lithographically defined geometry in the present work. 




\begin{table*}[ht]
\caption{\textbf{Qubit circuit parameters.} Data includes geometric properties such as the design type, the number of inductor turns $n$ and coil wire pitch $p$, as well as fitted qubit energies $E_{L}$, $E_{C}$, $E_{J}$ extracted from the data shown in Fig.~\ref{fig:all_spectra} in Appendix~\hyperref[spectra]{A}, the calculated zero point fluctuations of the phase $\varphi_\textrm{zpf}$, fitted coupling constants multiplied by the calculated matrix elements between states 0 and 1 at half flux, 
the measured values of $T_1$ also at half flux and the fitted flux noise amplitudes $A_\phi$.
}
\small\centering
\label{tab:summary}
\begin{tabular}{||c|c|c|c|c|c|c|c|c|c|c|c|c||} 
\hline
Qubit & design & $n$ & $p$ ($\mu$m) &$\tfrac{\kappa_\text{tot}}{2\pi}$ (MHz)&  $\tfrac{E_{L}}{h}$ (GHz) & $\tfrac{E_{C}}{h}$ (GHz) & $\tfrac{E_{J}}{h}$ (GHz) & $\varphi_\textrm{zpf}$ & $\tfrac{g_{C}\hat{n}}{2\pi}$ (MHz) & $\tfrac{g_{L}\hat{\phi}}{2\pi}$ (MHz) & $T_1$ ($\mu$s) & $A_\phi$ ($\mu \Phi_0$)\\

\hline
\hline

\textbf{A} & 2D & 74 & 0.4 &1.7& 0.618 & 2.75 & 8.55 & 1.73 & 15 & 0.1 & 1.5 & - \\
\hline
\textbf{B} & 2D & 74 & 0.4 &0.63& 0.620 & 3.15 & 5.92 & 1.78 & 63 & 140 & 2.38 & 317 \\
\hline
\textbf{C} & 2D & 74 & 0.4 &0.74& 0.619 & 3.25 & 5.41 & 1.80 & 69 & 100 & 3.29 & 338 \\
\hline
\textbf{D} & 2D & 74 & 0.4 &0.62& 0.620 & 3.83 & 3.05 & 1.88 & 41 & 210 & 1.81 & 787 \\
\hline
\textbf{E} & 2D & 125 & 0.3 &0.82& 0.205 & 2.97 & 4.89 & 2.32 & 6 & 2 & 9.62 & 673 \\
\hline
\textbf{F} & 2D & 125 & 0.3 &0.95& 0.215 & 3.40 & 1.99 & 2.42 & 90 & 7 & 2.25 & 646 \\
\hline
\textbf{G} & 3D & 70 & 0.3 &1.1& 0.78 & 0.50 & 3.15 & 1.06 & 17 & 0 & - & - \\
\hline
\textbf{H} & 3D & 25 & 0.25 &0.95& 10.70 & 0.54 & 9.00 & 0.56 & 98 & 0 & - & - \\
\hline

\end{tabular}
\end{table*}

\section{Model and Classification}
This section highlights the variety of qubits that can be achieved with this first generation of geometric rf-SQUID circuits and the different physics that can be observed. We present results from 8 different qubits that all derive from the circuit shown in Fig.~\ref{fig:fluxonium_modelling}(a) and are described by the Hamiltonian
\begin{equation}\label{eq:fluxonium_hamiltonian}
H_q = 4 E_\text{C} \hat{n}^2 + \frac{1}{2} E_\text{L} (\hat{\phi}-2\pi\Phi_\text{ext}/\Phi_0)^2 
-E_\text{J} \cos(\hat{\phi}),
\end{equation}
where $\hat{n}$ and $\hat{\phi}$ are the charge and phase operators,
$E_J=I_c\Phi_0/(2\pi)$ is the Josephson energy, $\Phi_\text{ext}$ is the external flux and $I_c$ the critical current of the single small Josephson junction, $\hat{n}$ is the charge and $\hat{\phi}$ the phase operator. The potential is given by the second and third term in Eq.~\ref{eq:fluxonium_hamiltonian} and the kinetic energy is given by the first term. It is important to note that Eq.~\ref{eq:fluxonium_hamiltonian} is the exact Hamiltonian 
of a Josephson junction shunted by an LC circuit
which is a faithful representation of the coil up to and above the first few qubit transitions \cite{Peruzzo2020}. The breakdown of the single mode lumped element model is described in Appendix~\hyperref[parasiticmode]{B}

What makes this circuit interesting is the fact that drastically different physics and qubit properties can be realized only by changing the relative magnitude of the three energies ($E_{C}$, $E_{L}$ and $E_{J}$) in the Hamiltonian. Figure \ref{fig:fluxonium_modelling}(c) presents a classification scheme that is valid around the flux frustration point $\Phi_\text{ext} = 0.5\, \Phi_0$ where qubits are often operated due to the first order protection from flux noise and  and the suppression of quasiparticles dissipation \cite{Pop2014}.

On the x-axis we have $E_{L}/E_{C} \propto 1/Z_{C}^2$ with $Z_{C} = \sqrt{L/C}$ the characteristic impedance of the qubit. This quantity affects the quadratic part of the Hamiltonian and will determine its overall steepness (gradient).
When it is small the phase is more delocalized since the wavefunctions of the ground and excited states bleed into more wells corresponding to higher zero point phase fluctuation values. As the qubit wavefunction spreads in phase the qubit transitions become less sensitive to flux noise.

On the y-axis we plot $E_J/E_C$ which is a measure of the relative depth of the sinusoidal wells. The inter-well coupling is exponentially dependent on $E_J/E_C$. Such coupling corresponds to the frequency of the qubit at the half flux quantum which can be made to dip down to 14 MHz \cite{Zhang2021} allowing for large $T_1$ according to Fermi's golden rule \cite{Schoelkopf2003}.
High  $E_J/E_C$ qubits are conventionally named 'heavy' due to their low kinetic energy.

From these distinctions four groups of qubits emerge which we classify according to their limit cases that may be modeled with different Hamiltonians and as indicated in Fig.~\ref{fig:fluxonium_modelling}(d):

\textit{The weakly anharmonic limit:} these light and low impedance circuits display a potential very similar to that of an harmonic oscillator but with very faint deformations due to the Josephson term. This results in a qubit state that resembles very closely that of an LC oscillator with a slight flux dispersion and a relatively small anharmonicity compared to its siblings. This limit might be interesting for a variety of applications including parametric amplification and for nonlinear oscillator based qubit implementations \cite{Grimm2020} but is not further studied in this work.

\textit{The flux qubit limit:} Qubits here have a strongly localized wavefunction due to $E_J$ and $E_L$ being higher than $E_C$. When $E_L$ is large (close to $E_J$) the system is similar to the flux qubits made out of three junctions in terms of spectrum and properties \cite{Yan2016}. In the case of $E_L$ close to  $E_C$ the physics is that of heavy fluxonium qubits where the high value of $E_J/E_C$ reduces the charge matrix element making state transfer challenging and typically requires involvement of higher order transitions \cite{Vool2018} or non adiabatic protocols \cite{Zhang2021} to prepare the first excited state at this flux value.

\textit{The fluxonium qubit limit:} Qubits in this limit have a high impedance and low inter-well tunneling. These were the first qubits to show phase delocalization by implementing an rf-SQUID with a superinductor \cite{Manucharyan2009}. It was shown that in the limit of $E_L/E_C<<1$ and $E_J/E_C \gtrapprox 2$ the lower energy levels and wavefunctions of this system are well described by a phase slip box Hamiltonian \cite{Koch2009,Ulrich2016}. The phase slip is the dual element of the Josephson junction and acts as a non-linear capacitor \cite{Mooij2006}.
In the limit of high $E_J/E_C$ however the approximation breaks down as many more levels come into play \cite{Koch2009}. Nevertheless, for intermediate impedance and $E_J/E_C\gg1$ we can identify a phase-qubit-like limit where the large superinductance generates the traditionally required current bias \cite{Martinis2003}, which is also accessible with geometric inductance \cite{Hassani2021} and represents an alternative strategy to control (and suppress) both flux dispersion and inter-well tunneling. 


\textit{The quasi-charge qubit limit:} These qubits are similar to the fluxonium qubit but due to low $E_L$ and low $E_J$ with respect to $E_C$ the wavefunction is able to spread further. In the case of very low $E_L/E_C$ \cite{Pechenezhskiy2020} the spread of the wavefunction will extend significantly beyond the two lowest potential wells making the qubit more insensitive to flux noise than its siblings. However, due to low $E_J/E_C$ this qubit has a relatively large matrix elements and hence the state is not well protected from relaxation.


In the following we experimentally access a range of 2-3 orders of magnitude in both $E_\text{L}/E_\text{C}$ and $E_J/E_C$. Figure \ref{fig:fluxonium_modelling}(d) shows the backed out parameters of qubits A - H reported in this work alongside representative rf-SQUID qubits based on kinetic inductance. 

\section{Qubit Designs}
\begin{figure}[t]
\centering
\includegraphics[width=\columnwidth]{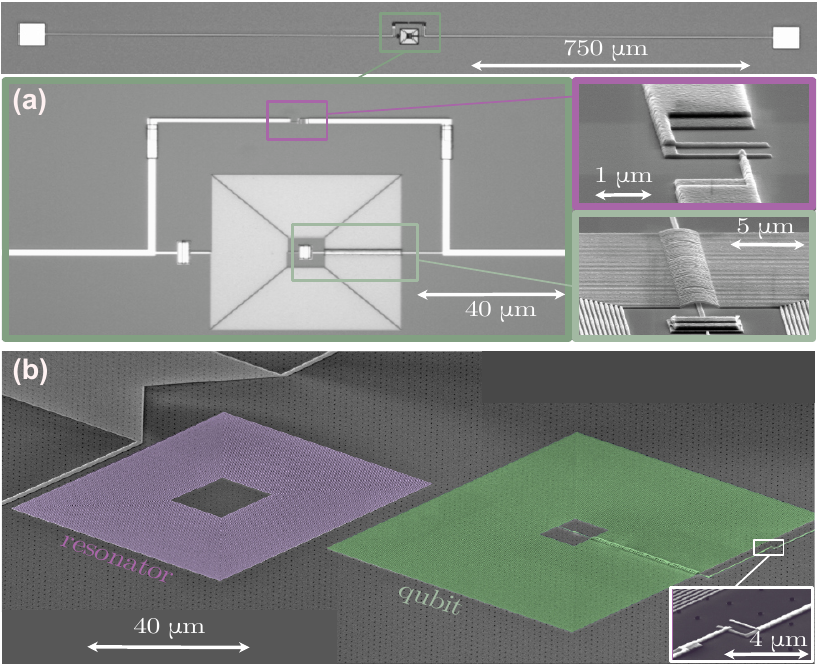}
\caption{\textbf{Qubit designs and fabrication.}  
(a) SEM of flux-type qubit G coupled to a 3D cavity via the coupling antenna (top) that was fabricated from aluminum on high resistivity silicon. Insets depict the planar coil inductor (green frame), with air bridges (bright green frame) and the Josephson junction (purple frame). 
(b) False-color SEM of quasi-charge qubit F (green) with single Josephson junction (inset) coupled to a resonator (purple), which is coupled to a shorted coplanar waveguide (top left). This device is fabricated from aluminum on a HF vapor etched silicon-on-insulator wafer.} 
\label{fig:fluxonium_design}
\end{figure}

Scanning electron micrographs (SEM) of a flux-type qubit (G) and a quasi-charge qubit (F) are shown in Fig.~\ref{fig:fluxonium_design} (a) and (b). The flux qubit is placed in a 3D cavity to which it couples via the large antenna that can be seen in the top inset. Such a coupling benefits from low dielectric loss and Purcell protection \cite{Pop2014} but it adds considerable capacitance making it incompatible with very light fluxonium devices. On the other hand, most devices in this study couple to a compact on-chip resonator and look like qubit (F) shown in Fig.~\ref{fig:fluxonium_design}(b). In this case both the qubit inductance (green) and resonator (purple) are made out of geometric superinductors, which results in an extremely compact footprint of about $60\, \mu$m$\, \times\, 120\, \mu$m for a full circuit QED system. 

The position of the two coils with respect to the coplanar waveguide coupler wire determines the coils' external coupling. The qubit coil is positioned symmetrically with respect to the coupler to minimize loss through this channel. The exact position was simulated with a finite element simulator, which predicted an extrinsic waveguide coupling limited $Q_\text{e}$ above one billion. Nevertheless, in practice the coupling allowed to address the qubit with sufficient field strength without an additional drive line and without measurable heating of the dilution refrigerator. Qubit operations such as a $\pi$ rotation were performed with pulse lengths down to 20 ns and where limited by instrument rise times.
The resonator coil on the other hand is asymmetrically positioned in order to obtain a $Q_{e}\approx2\times10^4$ to facilitate the readout. Coupling between the qubit and the resonator is obtained via the mutual inductance as well as the cross capacitance. For this work the qubits were placed in close proximity to the resonator (2-5 $\mu$m distance) with the magnetic qubit-resonator coupling reaching as high as 210 MHz for qubit D. 

When designing these qubits the inductance of the coil itself can be predicted analytically \cite{Mohan1999} with a 10-20 \% discrepancy with the measured values of the qubit inductance. If a comparably small amount of expected kinetic inductance is added \cite{Peruzzo2020} the predictions become accurate up to 5$\%$. As for the reproducibility of the inductive energy, qubits that had the same coil geometry (qubits A, B, C and D as well as qubits E and F) had a standard deviation of the inductance of only 0.2 \% and 0.7 \%. This is at least an order of magnitude more accurate than the typical Josephson energy variation of a simple single junction device without laser annealing and compatible with reproducibility requirements for symmetry protected or larger scale devices.

\section{Qubit Spectra}
So far rf-SQUID qubits have been coupled to a resonator either capacitively or inductively. The type of coupling is often trivially given by the qubit geometry, for example a shared inductance will give an inductive coupling while a large antenna placed in a 3D cavity will produce a capacitive coupling. While the flux qubits (qubits G and H) were capacitively coupled to a 3D cavity via an antenna, in the case of the other qubits explored in this paper neither coupling type can be completely excluded.
We model this situation by adding both interaction terms in the full system Hamiltonian 
\begin{equation}\label{eq:resonator_hamiltonian}
    H = H_q + \hbar \omega_r \hat{a}^{\dagger} \hat{a} -i \hbar g_C \hat{n} (\hat{a} - \hat{a}^{\dagger}) - \hbar g_L \hat{\phi} (\hat{a} + \hat{a}^{\dagger}),
\end{equation}
where $\omega_r$ is the resonator angular frequency and $\hat{a}$ ($\hat{a}^\dagger$) is the annihilation (creation) operator. The third and fourth terms are the capacitive and inductive coupling, where $g_{C}$ ($g_{L}$) is the capacitive (inductive) coupling constant.

\begin{figure*}[t]
\centering
\includegraphics[width=0.75\textwidth]{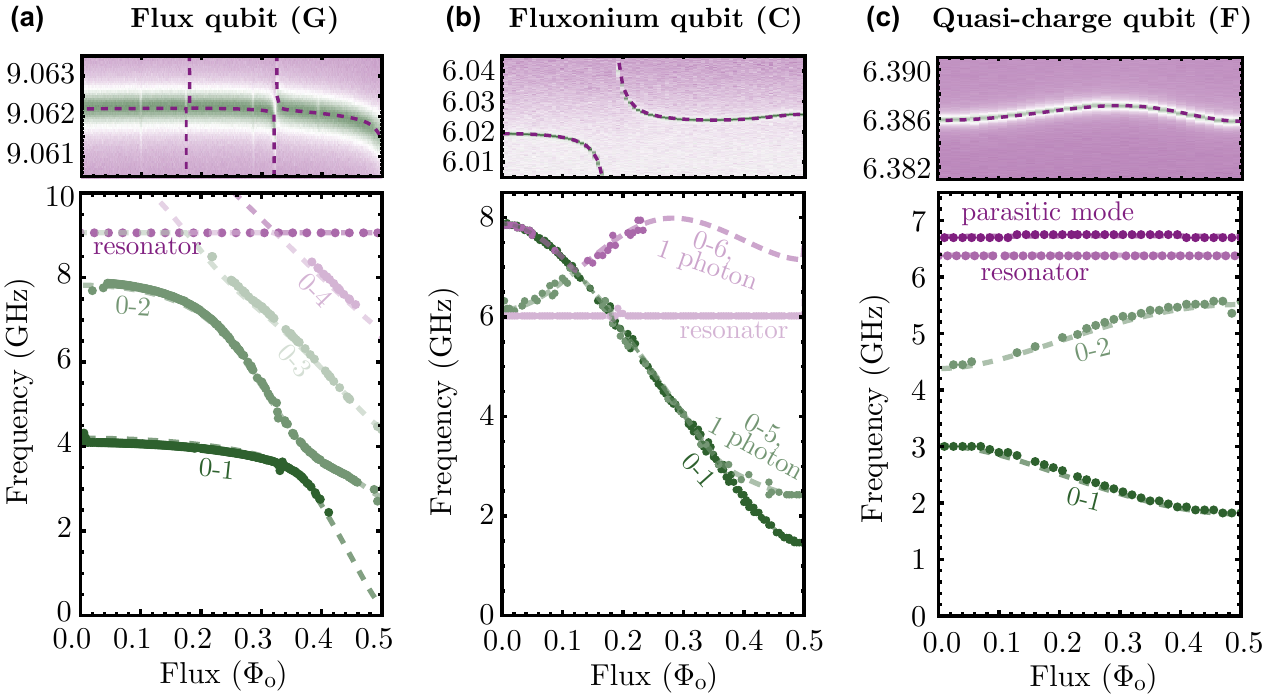}
\caption{\textbf{Flux dependent qubit and resonator spectra.}  
Measured resonator reflection amplitude (top, color scale) and two-tone spectroscopy (bottom, dots) of three representative qubits as a function of flux in panels (a) flux qubit, (b) fluxonium qubit, and (c) quasi-charge qubit.
The dotted lines are fits to the diagonalized spectrum of the full system Hamiltonian given by Eq.~\ref{eq:resonator_hamiltonian}. 
Only in the case of the quasi-charge qubit in panel (c) the qubit coil parasitic mode is significantly lowered due to a very large coil inductance and capacitance and becomes visible in the resonator measurement at 6.7 GHz. A quantitative model of the parasitic mode is presented in Appendix~\hyperref[parasiticmode]{B}.} 
\label{fig:fluxonium_spectrum}
\end{figure*}

Fitting fluxonium spectra and their coupling constants can be a challenge due to the multitude of parameters and their non-trivial effect on the energy levels. In an effort to make parameter fitting computationally tractable, the energy levels of the full Hamiltonian are computed using the in-place eigenvalue solvers for Hermitian matrices available in the Julia standard library. These solvers are highly performant, yielding eigenvalues for a $100 \times 100$ Hermitian matrix (5 photon states and 20 plasmon states) in less than 2 ms on a typical laptop (Intel Core i7-6700HQ CPU @ 2.60GHz x 8). The eigenvalue solver is then run repeatedly by a gradient-free local optimization routine \cite{Mogensen2018} until it converges on a 
set of parameters which best matches the data. 
Since the optimization routine is local, whether a trajectory converges on the correct solution
depends on the choice of initial parameters. 
The most efficient strategy is to first optimize the qubit parameters ($E_L$, $E_C$, $E_J$) to Eq.~\ref{eq:fluxonium_hamiltonian} and then adding the coupling constants ($g_L$ and $g_C$) and fitting to the full model while fixing values for qubit parameters. The code implementing the fitting of rf-SQUID spectra and coupling constants has been made available \cite{FluxoniumCode}. 

Figure~\ref{fig:fluxonium_spectrum}(a), (b) and (c) show spectra from three different qubits which reside in three of the categories presented in Fig.~\ref{fig:fluxonium_modelling}(c). 
Above each qubit spectrum the resonator dispersion versus flux is shown. These are reflective measurements taken with a vector network analyzer while the spectroscopy data is obtained with two tone spectroscopy. The resonators of qubits G and F have a relatively low variation in frequency due to a low coupling constant in the case of G and due to small flux dispersion for qubit F. On the other hand qubits A,B,C and D have high coupling constants and large flux dispersions (4-6 GHz) and therefore all display anti-crossings and large frequency variations in the dispersive shift. The quasi-charge qubit is the one with the smallest flux dispersion, this is due to the wavefunction residing in multiple wells at once. This qubit is comprised of one of the largest coils with the largest coil capacitance and inductance, lowering the frequency of the coil parasitic mode down into the accessible measurement range. Quantitative modeling of the parasitic mode coupling is presented in Appendix~\hyperref[parasiticmode]{B}.

A challenge in finding the correct coupling constants arose when trying to fit the flux dependent resonance shift, shown in the top panels of Fig.~\ref{fig:fluxonium_spectrum}. Out of the three predominant quantities (namely $g_{L}$ and $g_{C}$ and the resonator bare frequency $\omega_{r}$) only two are independent. This implies that there are a multitude of potential solutions. We narrowed this down to one solution by fixing the bare resonator frequency, which was measured separately at high probe power. 

The plots show excellent agreement with the diagonalized Hamiltonian. In addition to the bare qubit transitions many additional lines appear in the spectroscopy, these lines can easily be traced to being qubit transitions connected to a higher photon number. One of these transitions can be seen in Fig.~\ref{fig:fluxonium_spectrum}(b). All fitted qubit and coupling parameters are reported Table \ref{tab:summary}, and all measured and fitted spectra are presented in Appendix~\hyperref[spectra]{A}.

For qubits B, C and D - due to a relatively high transition frequency at the flux sweet spot and large qubit-resonator coupling - the dispersive shift is larger than the qubit linewidth which leads to photon number splitting of the qubit spectroscopy measurement similar to the one observed for transmon qubits \cite{Schuster2007a}. Measurements of photon number resolved fluxonium spectroscopy are shown in Appendix~\hyperref[numbersplitting]{C}. 

\section{Time domain analysis}
In this section we focus on the time domain results of a quasi-charge qubit (qubit F) and a fluxonium qubit (qubit E). Measured time-domain data from other qubits are summarized in Table \ref{tab:summary}. These two qubits were chosen as they have the same design, their only difference being the value of $E_{J}$. This difference however has strong implications for the qubit dynamics. Qubit E has a larger flux dispersion in the 0-1 qubit transition and hence is more susceptible to flux noise. However, due to the higher $E_{J}$ the qubit state is better protected from radiative decay due to a higher tunneling barrier. 

\begin{figure}[t]
\centering
\includegraphics[width=\columnwidth]{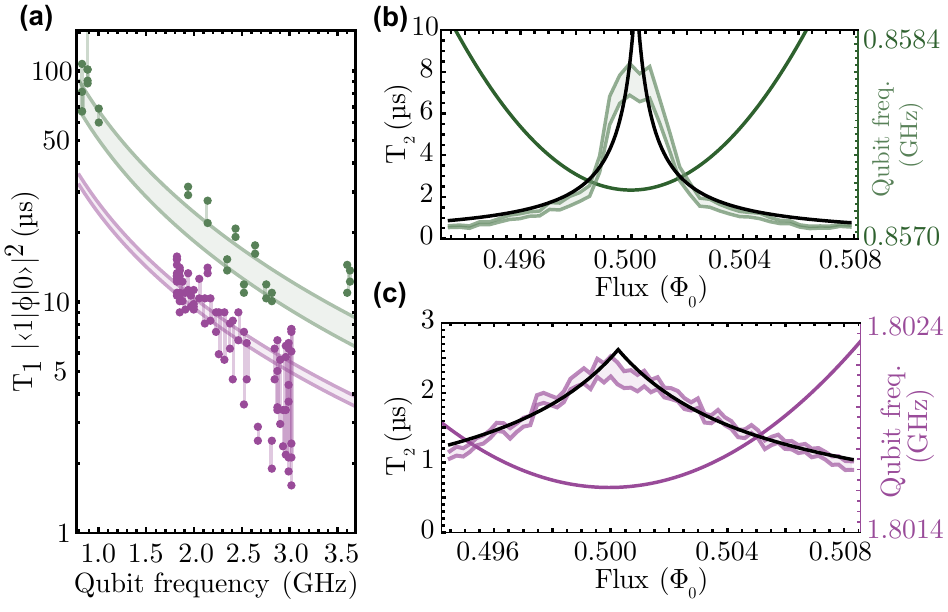}
\caption{\textbf{Time domain loss analysis.} 
(a), Measured energy relaxation time multiplied by the flux matrix element squared $T_1 |\mel{0}{\hat{\phi}}{1}|^2$ of the fluxonium qubit E (green) and the quasi-charge qubit F (purple) as a function of the external flux controlled qubit frequency. Fitted bands are to the dielectric loss model Eq.~\ref{eq:T1_vs_C}, with the fit parameter $Q_\text{diel}$
for qubits E and F. 
(b) and (c), Measured coherence time $T_2^\text{echo}$ as a function of external flux around half flux for qubits E and F. Bands represent the 90\% confidence interval of the spin echo sequence fit, the black line is a fit to the flux noise decoherence model Eq.~\ref{eq:T2_flux} and the purple (green) solid lines show the calculated qubit transition frequency (right axis).
} 
\label{fig:fluxonium_time}
\end{figure}

The effects of these characteristics are visible in Fig.~\ref{fig:fluxonium_time}(a), which shows the measured $T_1$ values multiplied by the absolute square of the phase matrix element as a function of the qubit transition frequency. Multiplying by the matrix element allows to compare the relaxation between qubits with different energies. The $T_1$ data was extracted placing the qubit in a mixed state via a saturation pulse and then measuring its decay with a dispersive readout. The extracted values are consistent with measurements done with a short (25-30 ns) $\pi$ pulse excitation at the flux sweet spot. 
The matrix elements were calculated numerically using the \emph{scqubits} python library \cite{scqubits}.

The behavior of the data agrees with a pure capacitive loss model, indicating that other loss mechanisms such as inductive loss or Purcell effect were not limiting. Even with the high coupling the Purcell limit to $T_1$ for these qubits is in the hundreds of $\mu$s due to the large detuning. We plot the data along side a temperature dependent capacitive loss model derived from Fermi's golden rule \cite{Schoelkopf2003}
\begin{equation}\label{eq:T1_vs_C}
\Gamma_1 = 1/T_1 = \frac{1}{(2e)^2} |\mel{0}{\hat{\phi}}{1}|^2 \hbar \omega_q^2\frac{C}{Q_\text{diel}} \coth (\frac{\hbar \omega}{2 k_B T}),
\end{equation}
where $\Gamma_1$ is the relaxation rate, $T_1$ is the relaxation time, $\mel{0}{\hat{\phi}}{1}$ is the phase matrix element, $\omega_q$ is the qubit's angular frequency, $C$ and $Q_\text{diel}$ are the total capacitance of the qubit and its quality factor and $T$ is the qubit temperature. 

The bands shown in Fig.~\ref{fig:fluxonium_time}(a) are fits to Eq.~\ref{eq:T1_vs_C} with a fitted $Q_\text{diel}$ of $(57 \pm 8)\times 10^3$ for the fluxoinum qubit and $(25 \pm 1) \times 10^3$ for the quasi-charge qubit. These values are similar to quality factors found for geometric superinductor resonators at single photon power on the same substrate (i.e. without handle wafer removal) \cite{Peruzzo2020} and on par with other fluxonium implementations \cite{Hazard2019}. The temperature is taken as 60 mK and 80 mK respectively, these numbers derive from the fit to the coherence data explained below. The values of $T_1$ of the fluxonium qubit are found to be consistently higher than the values for the quasi-charge qubit. This is in part due to a lower matrix element which stems from the larger tunneling barrier and a higher quality factor due to differences in fabrication. 

Compared to state of the art superconducting qubit devices the observed moderate values of $T_1$ in this first generation of geometric rf-SQUID qubits were obtained on an extremely small footprint with gap sizes as small as 250 nm which naturally leads to a higher sensitivity to two-level fluctuator induced loss. Better materials such as NbTiN or Ta and/or suitable surface treatments together with design optimizations that also includes removal of the handle wafer as in \cite{Peruzzo2020} should lead to a significantly longer $T_1$ in the near future.

Figure \ref{fig:fluxonium_time}(b) and (c) show $T_2^\text{echo}$ as a function of flux in the vicinity of the sweet spot for the same two qubits. The purple and green bands represents the coherence times and errors extracted from separate measurements. The black solid line represents the fit to a flux noise induced decoherence model with an upper bound of $ 2 T_1$ 
\begin{equation}\label{eq:T2_flux}
    \Gamma_2^\text{echo} = 1/ T_2^\text{echo} = \frac{\partial \omega_q}{\partial \Phi} \sqrt{A_\Phi} \gamma + \frac{1}{2 T_1} + \frac{1}{T_\phi},
\end{equation}
where $\Gamma_2^\text{echo}$ is the decoherence rate, $T_2^\text{echo}$ is the coherence time, $\Phi$ is flux, A$_\Phi$ is the noise amplitude of the flux noise spectral density $S_\Phi (\omega) = A_\Phi/\omega$, $\gamma$ is a constant which depends on the specific filtering function given by the chosen spin echo sequence and $T_\phi$ is a phenomenological constant added to account for photon shot noise due to the strong coupling to the resonator. This latter effect is exacerbated when the dispersive cavity shift 
is larger than the cavity linewidth 
\cite{Rigetti2012}, which is the case for qubits A-F.

The difference between the two types of qubits can be seen in how quickly the values of $T_2 ^\text{echo}$ decay as the flux is tuned away from the sweet spot. The dependence on the derivative of the transition frequency $\partial \omega_q / \partial \Phi$ stresses the advantage of making a quasi-charge qubit. For the two qubits presented in Fig.~\ref{fig:fluxonium_time}(b) and (c) we find $\sqrt{A_\Phi}$ to be 646\,$\mu \Phi_0$ and 673\,$\mu \Phi_0$ respectively. This is two orders of magnitude higher than what is measured in SQUIDs \cite{Braumuller2020} and other fluxonium devices \cite{Nguyen2019}. We attribute this to the long perimeter of the coil since the flux noise amplitude is likely due to magnetic spin defects at the surface and therefore perimeter dependent \cite{Braumuller2020}. The shot noise contribution was fitted to be 30.0 $\mu$s and 6.2 $\mu$s for qubits E and F respectively. These losses correspond to 0.03 and 0.006 photons leftover in the cavity, which in turn indicate a thermal bath of 80 mK and 60 mK. These numbers are on-par with similar implementations \cite{Yan2016} and could be further improved with better shielding.

  
\section{CONCLUSIONs and Outlook}
In this work we made an attempt to unify and classify the zoo of rf-SQUID qubits according to the physics rather than their physical implementation. This was guided by the new possibility of fabricating single superconducting wavefunction superinductors
that give access to a parameter range spanning three orders of magnitude from a highly localized to a strongly delocalized ground state phase wave function. 
As a result of our geometric inductance approach we observe 
simultaneous capacitive and inductive coupling between the qubit and the resonator and we provide a model and an algorithm that can efficiently fit the coupled rf-SQUID - resonator spectrum. We find couplings of the order of tens to hundreds of MHz achieved with very small coupling capacitance of 1 - 2 fF due to the small size and high impedance of both the qubit and the resonator - a feature that enables large capacitive coupling in the light fluxonium and quasi-charge regimes.

While the sensitivity to quasi-particle loss is expected to be much small compared to kinetic inductance qubits, the observed high flux noise amplitude is a potential disadvantage. This highlights the need for low flux dispersion by design, as achieved in the case of the measured quasi-charge qubit where $T_2$ is limited only by $T_1$ and the calculated shot noise limit. Other mitigation strategies include further miniaturization of the coil geometry in order to maximize the inductance per unit length as well as the use of materials with fewer magnetic surface defect states. On the positive side, this sensitivity points at other potential applications such as high precision quantum sensing of elementary spin systems.


The coherence and relaxation time will improve with more optimized design choices in future device generations but most importantly by increasing the quality factor of the inductor, for example by back-etching the handle wafer \cite{Peruzzo2020}, which would incidentally also allow to reach even lower values of $E_\text{L}/E_\text{C}$. The resulting further enhanced zero point phase fluctuations are a prerequisite towards the realization of degenerate ground state qubits, where the full protection requires 
Hamiltonian engineering with carefully maintained circuit symmetries, a characteristics of top-down fabricated circuit elements.

We believe that the demonstrated design flexibility, the low chip-to chip variance of the capacitive and inductive energy of $< 1$\,\%, the simplicity of predicting the physics of 2-terminal devices, the ability to reach very high inductance values and the new capability of strong magnetic coupling to resonators, waveguides or other qubits make geometric superinductor qubits an interesting and complementary ingredient towards realizing hardware protected qubits in new parameter regimes in the near future. 


The data and code used to produce the figures in this manuscript will be made available at Zenodo. 

\section*{Acknowledgments}
The authors thank W. Hughes for analytic and numerical modeling during the early stages of this work, J. Koch for discussions and support with the \emph{scqubits} package, R. Sett P. Zielinski and L. Drmic for software development, G. Katsaros for equipment support, as well as the MIBA workshop and the IST nanofabrication facility. We thank I. Pop, S. Deleglise and E. Flurin for discussions. This work was supported by a NOMIS foundation research grant, the Austrian Science Fund (FWF) through BeyondC (F7105), and IST Austria. M.P. is the recipient of a P\"ottinger scholarship at IST Austria. E.R. is the recipient of a DOC fellowship of the Austrian Academy of Sciences at IST Austria.


\section*{Appendix A: spectra of all qubits}\label{spectra}
\begin{figure*}[t]
\centering
\includegraphics[width=0.8\textwidth]{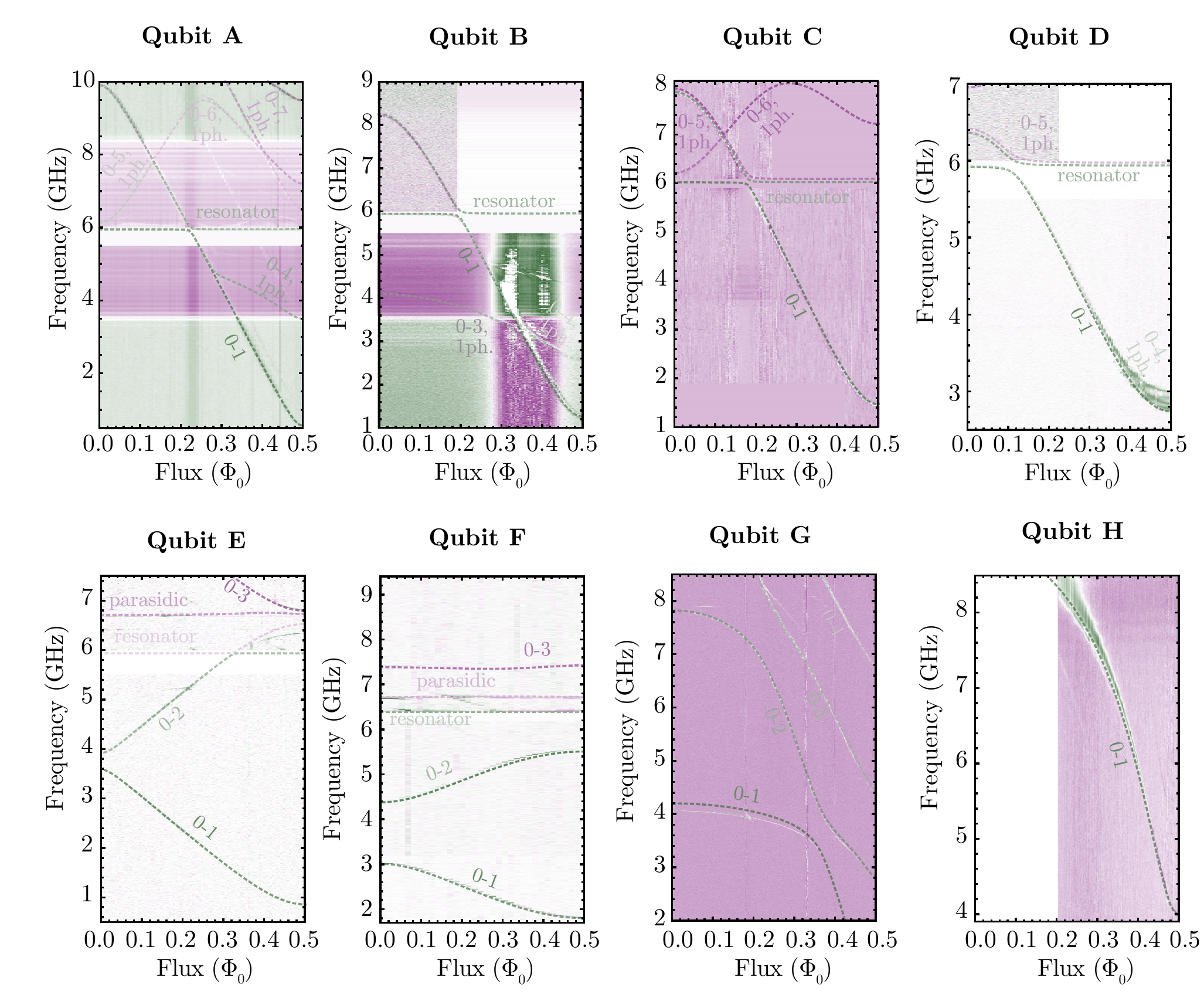}
\caption{\textbf{Two-tone spectroscopy of all measured qubits.} The labeled lines are obtained from a numerical optimization algorithm of the qubit and coupling parameters to maximize agreement with the full rf-SQUID and resonator model given in Eq.~\ref{eq:resonator_hamiltonian}. In the case of qubits E and F the observed parasitic qubit mode was added as a second coupled resonator mode to the Hamiltonian.} 
\label{fig:all_spectra}
\end{figure*}

The spectroscopy data of all qubits can be found in Fig. \ref{fig:all_spectra}. These show a clean spectrum where most lines can be attributed to qubit or dressed states. The parameters corresponding to the fits are reported in Table \ref{tab:summary} alongside geometric parameters of the samples and the time domain measurement results that were conducted.

\section*{Appendix B: coupling to the parasitic mode}\label{parasiticmode}
\begin{figure}[t]
\centering
\includegraphics[width=0.9\columnwidth]{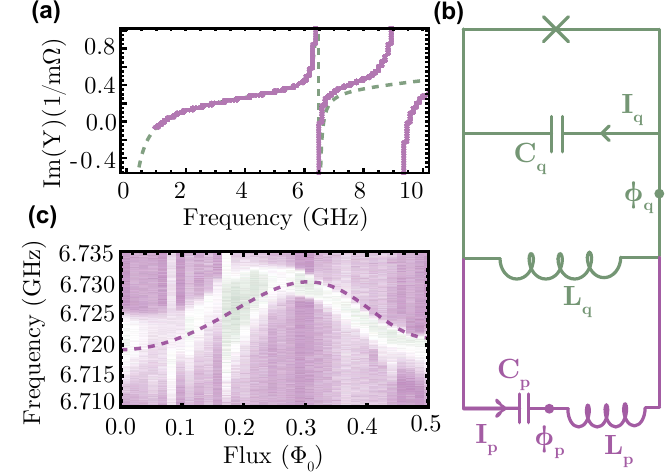}
\caption{\textbf{Qubit parasitic mode.} 
(a) Simulated admittance of the coil used in qubits E and F. Purple dots represent the simulation, green dashed line is a fit to the circuit shown in panel (b). Fitted parameters are $C_{q} = 4.8$ fF, $L_{q} = 530$ nH, $C_{p} = 0.47$ fF, $L_{p} = 1.3~\mu$H. The simulated data shows an additional pole appearing at higher frequency that was not identified in the qubit spectrum. 
(b) A phenomenological circuit model for the high frequency response of the coil inductor that shows very good agreement with the simulated admittance up around 8 GHz. The green part of the circuit represents the qubit while the purple part models the parasitic mode. 
(c) Two tone spectroscopy data of the parasitic mode of the qubit. The dashed line is obtained by solving the full Hamiltonian containing the qubit, the resonator mode and the parasitic mode. Here the coupling to the parasitic mode was taken to be 0.84 GHz as correctly predicted by Eq.~\ref{eq:parasitic_mode_coupling}.
} 
\label{fig:parasitic_mode}
\end{figure}

The planar coil used as a superinductor is a distributed element circuit. At low frequencies it can be described by a simple lumped element model, i.e. a parallel LC circuit where the fundamental mode forms the first transition frequency of the qubit. The second mode of the coil inductor acts as a resonator mode that couples strongly to this qubit mode. This is in contrast to other fluxonium qubit implementations where the parasitic mode is given by the first resonance of the superinductor \cite{Hazard2019,Rastelli2015,Viola2015}.
We use finite element simulations of the planar coil inductor used for qubits E and F to predict the frequency of this mode as shown in Fig.~\ref{fig:parasitic_mode}(a). At a frequency of about 6.5 GHz the admittance shows an additional pole compared to the simple LC model. We model this by adding in parallel to the original LC circuit an extra inductance and capacitance in series as shown in Fig.~\ref{fig:parasitic_mode}(b) and agrees well with the simulated values. In the future this mode could therefore potentially be used as a built in read out resonator to further simplify the circuit design.

The Lagrangian for this circuit is
\begin{equation}\label{eq:parasidic_hamiltonian}
\begin{aligned}
    \mathcal{L} = C_q (\dot{\phi}_q)^2+C_p (\dot{\phi}_q-\dot{\phi}_p)^2+\frac{1}{2} E_{L,q} \phi^2_q \\ 
    +\frac{1}{2} E_{L,p}\phi^2_p -E_{J} \cos{\phi_q},
\end{aligned}
\end{equation}
where $C_{q}$ ($C_{p}$) and $\phi_q$ ($\phi_p$) represent the capacitance and phase variable of the qubit (parasitic mode) and $E_{L,q}$ and $E_{L,p}$ represent the inductive energies of the qubit and parasitic mode, corresponding to $L_{q}$ and $L_{p}$ as shown in Fig.~\ref{fig:parasitic_mode}(b).
Next the relation between the node voltages and currents of the circuit must be identified. 
This enables the replacement of the node voltages $\dot{\phi}_{q}$ and $\dot{\phi}_{p}$ with their respective canonical conjugates $Q_{q}=\pder[I_{q}]{t}$ and $Q_p=\pder[I_{p}]{t}$. Using Kirchhoff's laws one can write the following equations in the Fourier domain
\begin{equation}\label{eq:relations}
\begin{cases} V_{q}-V_{p}=\frac{I_{p}}{C_{p}S} \\ \\V_{q}=\frac{I_{q}-I_{p}}{C_{p}S}  \end{cases},
\end{equation}
where $S$ is the complex frequency, which corresponds to a differentiation operator in the time domain $\frac{d}{dt}$, while $\frac{1}{S}$ is an integration operator$\int dt$. 

Using these relations and the fact that $V_{q}=\dot{\phi}_{q}$ and $V_{p}=\dot{\phi}_p$ we can rewrite Eq.~\ref{eq:relations} as
\begin{equation}\label{eq:relations_2}
\begin{cases}\dot{\phi_{p}}=\frac{Q_{q}-Q_{p}}{C_{q}}-\frac{Q_{p}}{C_{p}} \\ \\\dot{\phi_{q}}=\frac{Q_{q}-Q_{p}}{C_{q}}. \end{cases}
\end{equation}
Replacing Eq.~\ref{eq:relations_2} into Eq.~\ref{eq:parasidic_hamiltonian} gives rise to the final form of the total Hamiltonian
\begin{equation}\label{eq:parasidic_hamiltonian_final}
    \begin{aligned}
H = H_\text{parasitic}+H_\text{qubit}+H_\text{coupling}= \\
\quad\frac{\hat{Q}^2_{p}}{(C_{p}^{-1} + C_{q}^{-1})^{-1}}+\frac{1}{2}E_{L,p}\hat{\phi}^2_{p}+\frac{1}{2}E_{L,q}\hat{\phi}^2_{q}
-E_{J}\cos{\hat{\phi}_{q}} \\
+\frac{\hat{Q}^2_{q}}{C_{q}}+\frac{2\hat{Q}_{q} \hat{Q}_{p}}{C_{q}}
\end{aligned}
\end{equation}
where we quantize $\hat{\phi}$ and $\hat{Q}$.
From Eq.~\ref{eq:parasidic_hamiltonian_final} it is possible to identify the frequency of the parasitic mode as
\begin{equation}\label{eq:parasitic_mode_freq}
    \omega_{p} = \sqrt{\frac{1}{L_{p} (C_{p}^{-1} + C_{q}^{-1})^{-1}}}.
\end{equation}

Rewriting the coupling term of Eq.~\ref{eq:parasidic_hamiltonian_final} in the second quantization formalism results in
\begin{equation}
	\begin{aligned}
   H_\text{coupling}=\frac{2 \hat{Q}_q \hat{Q}_p}{C_q} =\frac{2 \hat{Q}_q}{C_q}(\hat{a}+\hat{a}^\dagger)\sqrt{\frac{\hbar}{2 Z_p}}= \\
   \frac{2}{C_q}\sqrt{\frac{\hbar}{2 Z_p}}(\hat{a}+\hat{a}^\dagger)\hat{Q}_q,
	\end{aligned}  
\end{equation}
which identifies the coupling strength as
\begin{equation}\label{eq:parasitic_mode_coupling}
    g_{p} = \frac{4e}{C_{q}} \sqrt{\frac{\hbar \omega_{p}(C_{p}^{-1} + C_{q}^{-1})^{-1}}{2}}.
\end{equation}
 	
By fitting the coil admittance found in Fig.~\ref{fig:parasitic_mode}(a) we find the parasitic inductance to be $L_{p} = 1.28~\mu$H and the parasitic capacitance to be $C_{p} = 0.47$ fF. By inserting these values into Eq.~\ref{eq:parasitic_mode_freq} alongside the fitted qubit parameters the frequency of the parasitic mode is expected at 6.74 GHz, very close to the measured value seen in Fig. \ref{fig:parasitic_mode}(c) and the coupling is expected to be $g_{p} =0.84$ GHz. The fit line in Fig.~\ref{fig:parasitic_mode}(c) is obtained by diagonalizing the Hamiltonian of the whole system consisting of the qubit, the resonator and the parasitic mode where the parasitic mode is added as an additional resonator coupled to the qubit. For the fit the frequency of the parasitic mode was taken to be 6.73 GHz while the coupling was taken from Eq.~\ref{eq:parasitic_mode_coupling} with values based on the simulated admittance.

\begin{figure}[b!]
    \centering
    \includegraphics[width=\columnwidth]{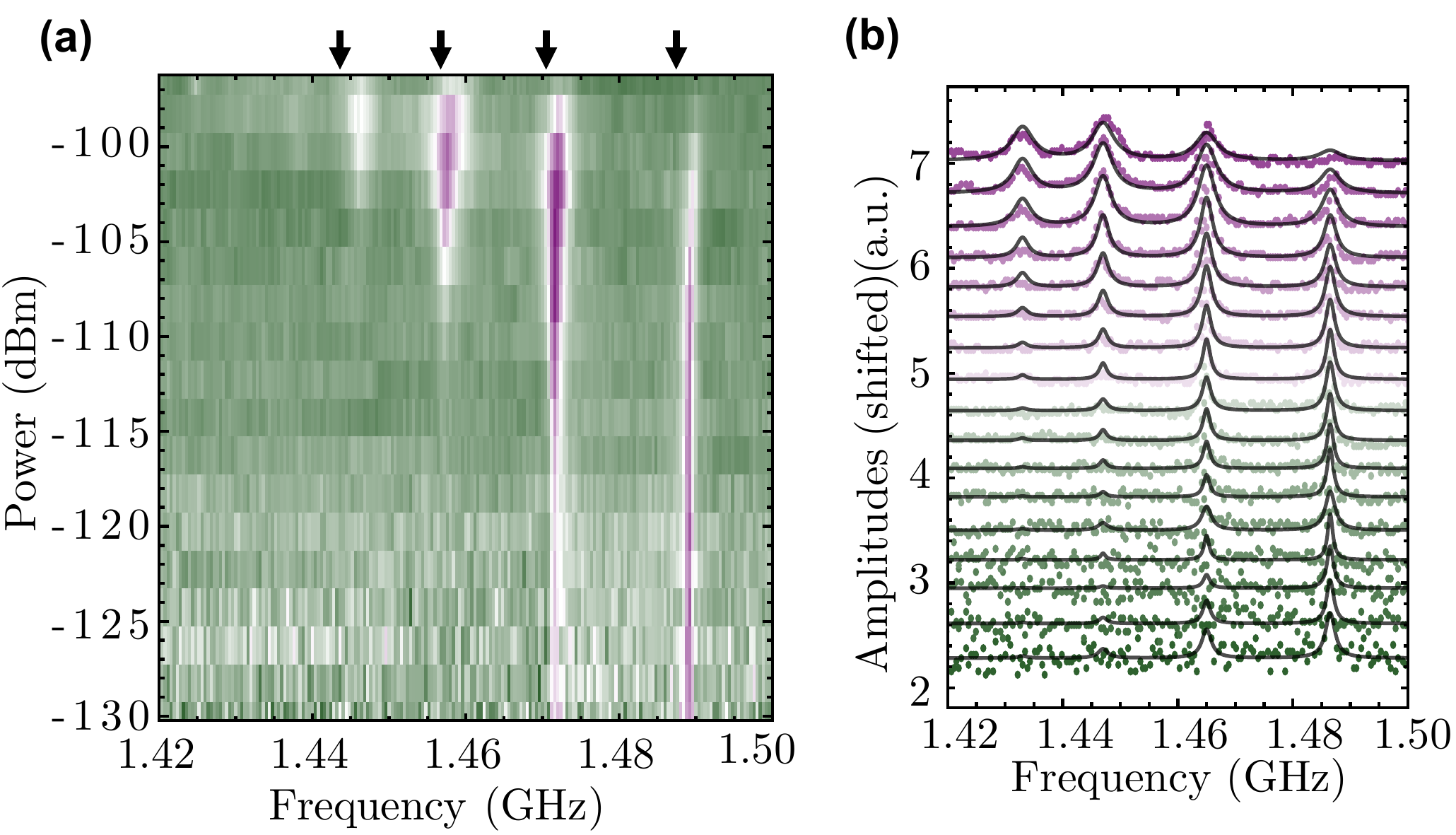}
    \caption{\textbf{Measured photon number splitting.}
    (a) Spectroscopy of qubit C as a function of measurement power. Arrows indicate the solution of the full Hamiltonian. 
    (b) Individual measurement traces from panel (a) plotted with an offset for visibility. The black lines correspond to fits to a Poissonian weighted sum of Lorentzians. }\label{FigNumSplit}
\end{figure}
    
\section*{Appendix C: Fluxonium photon number splitting}\label{numbersplitting}
Due to large coupling strengths of qubits B, C and D they all showed some form of resolved photon number splitting at both the sweet spots of the first qubit transition. Figure \ref{FigNumSplit} shows the effect measured in qubit C as a function of resonator measurement power. The resonators for these qubits are all found around 6 GHz, specifically 6.03 GHz in the case qubit C. Figure \ref{FigNumSplit}(a) shows a 2D plot of the measurement. We observe that the separation between the photon number resolved spectroscopy peaks 
is in agreement with the full Hamiltonian. Figure \ref{FigNumSplit}(b) shows the individual traces from panel (a) offset for better visibility and fitted to a series of Lorenzians whose amplitudes are fixed by a Poisson distribution. In this measurement 
we observe a sizable single photon excitation probability of the resonator indicating that an improved shielding is necessary to avoid excess qubit dephasing. Such improvements were implemented for later measurements of qubits E and F.

\bibliography{Bib}

\end{document}